\documentclass[prl,twocolumn]{revtex4} 
\usepackage{graphicx,amsmath}

\begin{document}

\newcommand{\ket}[1]{\mbox{$|\!#1\;\!\rangle$}}
\newcommand{\aver}[1]{\mbox{$<\!#1\!\!>$}}
\def\ua{\uparrow}
\def\da{\downarrow}

\title{Locking electron spins into magnetic resonance by electron-nuclear feedback}

\author{I. T. Vink}
\affiliation{Kavli Institute of Nanoscience, Delft University of
Technology,\\
PO Box 5046, 2600 GA Delft, The Netherlands}
\author{K. C. Nowack}
\affiliation{Kavli Institute of Nanoscience, Delft University of
Technology,\\
PO Box 5046, 2600 GA Delft, The Netherlands}
\author{F. H. L. Koppens}
\affiliation{Kavli Institute of Nanoscience, Delft University of
Technology,\\
PO Box 5046, 2600 GA Delft, The Netherlands}
\author{J. Danon}
\affiliation{Kavli Institute of Nanoscience, Delft University of
Technology,\\
PO Box 5046, 2600 GA Delft, The Netherlands}
\author{Yu. V. Nazarov}
\affiliation{Kavli Institute of Nanoscience, Delft University of
Technology,\\
PO Box 5046, 2600 GA Delft, The Netherlands}
\author{L. M. K. Vandersypen}
\affiliation{Kavli Institute of Nanoscience, Delft University of
Technology,\\
PO Box 5046, 2600 GA Delft, The Netherlands}

\date{\today}

\maketitle

{\bf 
The main obstacle to coherent control of two-level quantum systems is their coupling to an uncontrolled environment~\cite{zurek}. For electron spins in III-V quantum dots, the random environment is mostly given by the nuclear spins in the quantum dot host material; they collectively act on the electron spin through the hyperfine interaction, much like a random magnetic field ~\cite{khaetskii_prl,merkulov_prl,sigi_prb,petta_science,koppens_prl08,greilich_science06,hanson_rmp}. Here we show that the same hyperfine interaction can be harnessed such that partial control of the normally uncontrolled environment becomes possible. In particular, we observe that the electron spin resonance frequency remains locked to the frequency of an applied microwave magnetic field, even when the external magnetic field or the excitation frequency are changed. The nuclear field thereby adjusts itself such that the electron spin resonance condition remains satisfied. General theoretical arguments indicate that this spin resonance locking is accompanied by a significant reduction of the randomness in the nuclear field.}


In thermodynamic equilibrium, the nuclear spins in the quantum dot host material are randomly oriented, even at dilution refrigerator temperatures and in magnetic fields of a few Tesla. An electron spin confined in the quantum dot interacts via the hyperfine coupling with $N \sim 10^6$ nuclear spins and as a result experiences a random nuclear field $B_N$. This random nuclear field is sampled from a distribution with a root mean square width $\propto A/g\mu_B\sqrt{N}$, where $g$ is the electron $g$-factor, $\mu_B$ the Bohr magneton and $A$ the hyperfine coupling constant ($\approx 135 \mu$eV in GaAs). Measurements typically give a width of $\sim 1$ mT. As a result, we lose track of the phase of a freely evolving electron spin within a time $T_2^*$ of a few tens of nanoseconds~\cite{khaetskii_prl,merkulov_prl,petta_science,koppens_prl08,greilich_science06}. Similarly, when the spin evolves under an oscillating driving field, the nuclear field leads to a random offset in the resonance condition which has a comparable amplitude to presently achievable driving fields. This results in poorly controlled spin rotations~\cite{koppens_nature}. 

It is therefore of great importance to develop the ability to control and manipulate the nuclear field with great precision. In particular, it would be highly desirable to set the nuclear field to a narrow distribution of values at the start of every experiment~\cite{burkard_prb,klauser,giedke,stepanenko}. This would immediately reduce the rapid dephasing, and the electron spin would loose phase coherence only from the slow subsequent evolution of the nuclear field, giving a predicted spin coherence time of $1-10 \mu$s~\cite{witzel_prb,coish_prb}. Such narrowing has been achieved in an ensemble of self-assembled quantum dots by synchronizing the precessing spins with a series of laser pulses~\cite{greilich_science07}. More recently, the spread of the difference in nuclear fields in two neighbouring quantum dots was reduced via a gate voltage controlled pumping cycle, giving a 70-fold increase in the $T_2^*$ for states in the two-electron $m_z=0$ subspace~\cite{reilly_zamboni}.

Here we exploit electron-nuclear feedback in order to control and manipulate the nuclear fields in two coupled quantum dots during continuous wave (CW) driving of the electron spins in the dots. We observe that the nuclear field adjusts itself such that the electron spins remain in resonance with a fixed driving frequency even when we sweep the external magnetic field away from the nominal resonance condition. Similarly, the electron spin resonance frequency remains locked to the excitation frequency when the excitation frequency is swept back and forth. These distinctive features set our observations apart from the many previous observations of dynamic nuclear spin polarization in quantum dots, both in transport~\cite{ono_prl,koppens_science,reilly_prl,baugh_prl,foletti_arx} and optical measurements~\cite{tartakovskii,maletinsky_prb,braun_prb}. We investigate the origin of this feedback by studying its dependence on the amplitudes of the applied $ac$ magnetic and electric fields and on the sweep rates. We show theoretically that the spin resonance locking must be accompanied by a narrowing of the nuclear field distribution, in the present experiment by more than a factor of 10.

\begin{figure}[!b]
\includegraphics[width=8.5cm]{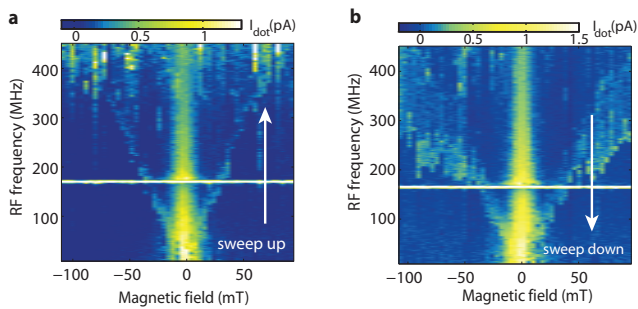}
\caption{Electron spin resonance locking during frequency sweeps. \textbf{a}, Current through the double dot (colorscale) subject to CW magnetic excitation, when sweeping the frequency up at fixed magnetic fields. The bright fork indicates the position of the ESR condition. \textbf{b}, Similar to \textbf{a} but sweeping the frequency down. The ESR frequency remains locked to the excitation frequency when the excitation frequency is swept past the nominal resonance condition. 
}
\label{fig:nuclear_polarization}
\end{figure}

The measurements are performed on an electrostatically defined double quantum dot tuned to the Pauli spin blockade regime~\cite{ono_science}, with effectively one excess electron on each dot (the actual electron number is small but unknown). When the two electrons have parallel spins, the electron flow through the dots is blocked. When one of the spins is flipped, the spin blockade is lifted and electrons flow through the two dots until the system returns to a state with parallel spins on the two dots. As previously demonstrated~\cite{koppens_nature}, it is possible to flip the dot spins via magnetic resonance, by $ac$ excitation of an on-chip wire which generates an oscillating magnetic field at the dots: when the excitation frequency, $f$, matches the electron spin resonance (ESR) frequency, $|g|\mu_{B}B_{0}/h$, a finite current flows through the device. Here $h$ is Planck's constant, and $B_0$ the external magnetic field. In addition, current can flow at zero magnetic field, where the electron spins can flip-flop with the nuclear spins in the substrate~\cite{koppens_science}. We use this zero-field feature to determine and adjust for small magnetic field offsets present in our setup. The zero-field peak and the ESR response are seen in current measurements under CW excitation with increasing excitation frequency at fixed magnetic fields (Fig.~\ref{fig:nuclear_polarization}a, similar to the data published in Ref.~\cite{koppens_nature}, and taken on the same device but in a different cooldown.

Surprisingly, when we reverse the sweep direction, a distinctly different behavior is observed over a wide range of dot settings (see Supplementary Information for details of the tuning parameters). Current starts flowing when the driving frequency hits the spin resonance frequency but \textsl{remains} high even as the frequency is swept well below the nominal resonance condition (Fig. \ref{fig:nuclear_polarization}b. The fact that the current remains high implies that the electron spin is still on resonance with the excitation frequency, and that an effective field, $B_\text{eff}$, counteracts the external magnetic field $B_0$: $hf=|g|\mu_{B}(B_{0}+B_\text{eff})$. From the fact that the current is strongly reduced when we simultaneously excite any of the three nuclear spin species in the substrate (data not shown), we conclude that this effective field is created by dynamical nuclear spin polarization. This nuclear field builds up exactly at the right rate in order to keep the electron spin in resonance with the changing driving frequency, which implies there is a built-in electron-nuclear feedback mechanism.

\begin{figure}[!t]
\includegraphics[width=7cm]{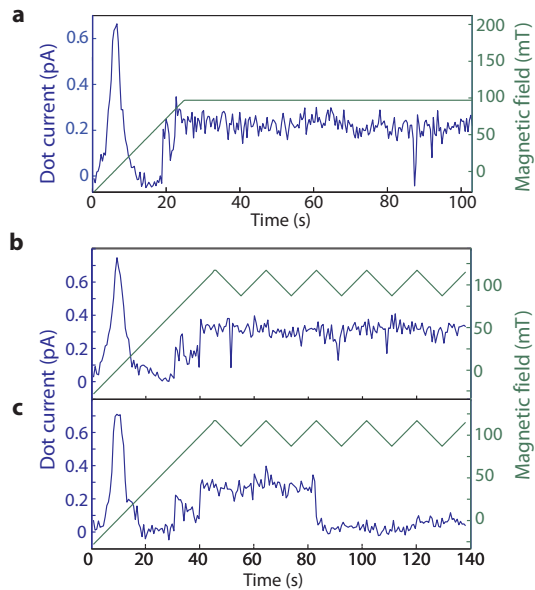}
\caption{Electron spin resonance locking during magnetic field sweeps. \textbf{a}, Current through the double dot as a function of time, while the magnetic field is first ramped up (right axis) and subsequently held fixed, under CW excitation ($f=400$ MHz). \textbf{b-c}, Two current traces similar to \textbf{a}, but after the magnetic field is ramped up, it is repeatedly swept down and back up over a 30 mT range (right axis). After the ESR condition is first met, the electron spin remains locked into magnetic resonance for up to two minutes, even though the resonance condition is shifted back and forth.}
\label{fig:sawtooth_wait}
\end{figure}

\begin{figure}[!t]
\includegraphics[width=8.5cm]{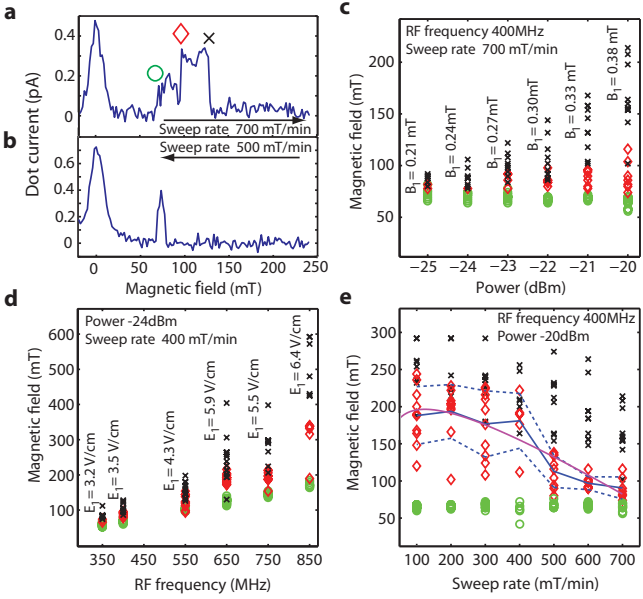}
\caption{ESR locking dependence on excitation power, frequency and sweep rate. \textbf{a}, Current through the double dot as the magnetic field is swept up ($f=400$ MHz). 
\textbf{b}, Similar to \textbf{a} but now $B_0$ is swept down. No dragging effects are observed; the narrow peak gives the position of the nominal resonant field. \textbf{c}, Scatter plot of the switching fields as indicated by the symbols in \textbf{a}, as a function of the power applied to the on-chip wire. The corresponding resonant magnetic field amplitude $B_1$ at the dot is given as well. \textbf{d}, Scatter plot similar to \textbf{c}, as a function of $f$. The estimated electric field amplitude $E_1$ increases with $f$, and is shown in the figure. \textbf{e}, Scatter plot similar to \textbf{c} as a function of magnetic field sweep rate. Blue lines: average and standard deviation of the magnetic fields where the second current jump is observed. Purple curve: fit of these average values with a theoretical model (see Supplementary Information). We note that there is no build-up of $B_N$ in the limit of zero sweep rate, so the predicted switching field first increases with sweep rate, before decreasing.}
\label{fig:sweeps}
\end{figure}

Similar dragging of the resonance is observed when sweeping the magnetic field for a fixed excitation frequency. In Fig.~\ref{fig:sawtooth_wait}a the magnetic field is swept from -33 mT to 97 mT (right vertical axis) in about 25 seconds. We first see the zero-field peak, as expected, and next the current jumps up around $B_0=67$ mT, which is slightly below the nominal resonance condition ($f = 400$ MHz, $|g|=0.36$). The current remains high as the field is swept further to 97 mT, which is well outside the ESR linewidth in the absence of feedback (see Fig.~\ref{fig:sweeps}b). Similar to the case of the frequency sweeps, a nuclear field builds up exactly in such a way as to maintain the ESR frequency locked to the excitation frequency. When we subsequently keep the field fixed at 97 mT, we observe that the electron spin can remain locked into magnetic resonance for well over a minute. 

It is also possible to drag the nuclear field back and forth under fixed-frequency excitation. In Figs.~\ref{fig:sawtooth_wait}b and ~\ref{fig:sawtooth_wait}c, $B_0$ is ramped up from -33 mT to 117 mT, and is subsequently swept back and forth between 117 mT and 87 mT in a triangular pattern. The current again jumps up as we sweep through resonance and subsequently remains high independent of the sweep direction. In Fig. \ref{fig:sawtooth_wait}c the resonance is lost after approximately 1 minute, whereas in Fig. \ref{fig:sawtooth_wait}b the spin remains locked on resonance during the entire experiment (about 2 minutes).

These remarkable observations of spin resonance locking due to electron-nuclear feedback are characterized by a number of common features.
First, the current jumps up abruptly, in many cases in less than a few 100 ms, at a field value that varies over 10-30 mT around the nominal resonance condition (see the green circles in Fig.~\ref{fig:sweeps} below). This is a further indication that the system is actively pulled into resonance -- without feedback a current peak with smooth flanks and a width of a few mT is expected~\cite{koppens_jap}. 
Second, the resonance dragging generally occurs only for fields larger than the nominal resonant field, or for frequencies lower than the nominal resonance frequency. This is opposite to the case of the usual Overhauser effect, as discussed further below.
Third, the initial current jump is usually followed by a second current jump, before the current drops back to zero. A possible explanation for this double step is that the first current plateau corresponds to a situation where both electrons are on-resonance, and that only one electron remains on resonance after the second jump (see Supplementary Information for a discussion of the current levels). When the resonance is lost in this last dot too, the current returns to zero. 

\begin{figure}[!t]
\includegraphics[width=6.5cm]{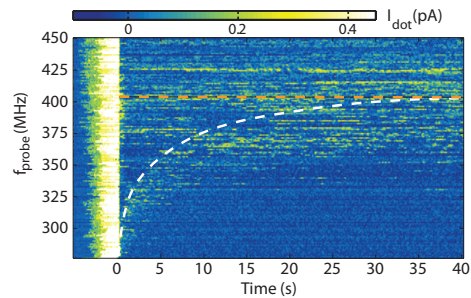}
\caption{Pump-probe measurement of the relaxation of the nuclear spin polarization. At a fixed magnetic field of $B_0=80$ mT, we apply CW excitation ($P=-13$ dBm) sweeping the frequency from 500 MHz to 276 MHz at $43$ MHz/s, and dragging the nuclear field along (pump phase). Next we turn off the CW excitation and apply 140 ns microwave bursts every 2 $\mu$s at frequency $f_\text{probe}$ throughout a 40 s probe phase. This pump-probe cycle is repeated for different probe frequencies, 277 MHz $\le f_\text{probe} \le$ 450 MHz (see vertical axis). The horizontal axis indicates the time $t$ into the probe phase; the data for $t<0$ correspond to the pump phase. 
In the pump phase, the current (plotted in colorscale) jumps up twice, reaching the highest current plateau (traces where the resonance is lost by the end of the pump phase are left out). When the frequency is switched to $f_\text{probe}$ at $t=0$, the current drops to zero since the excitation is now off-resonance. As the nuclear spin polarization relaxes, the resonance condition $\left|g\right|\mu_{B}(B_{0}+B_{N}(t))=hf_\text{probe}$ will be fulfilled at some point in time at which the current sets on again. Varying $f_\text{probe}$ reveals then the nuclear spin relaxation as indicated by the white dashed line (guide to the eye) marking the onset of the current. Even though the excitation is applied only in bursts, the electron spin nevertheless remains locked into resonance in some cases, stalling the nuclear spin relaxation. The orange dashed line marks an additional signal at the nominal resonance frequency already present from the start of the probe phase.}
\label{fig:pump_probe}
\end{figure}

This interpretation of the double current step is supported by pump-probe measurements shown in Fig.~\ref{fig:pump_probe}. Starting from the second current plateau with $B_0=80$ mT and $f=276$ MHz, we switch off the CW excitation and probe the position of the ESR frequency as the nuclear field returns to equilibrium (we use short bursts for probing in order to minimize feedback during the probe phase). We see that the ESR frequency returns to its nominal value, slightly above 400 MHz, within 20 seconds, corresponding to the relaxation time of the local nuclear spin polarization. 
This signal must originate from the dot that is still locked into magnetic resonance at the end of the pump phase. In addition, we see a response at the nominal resonance frequency already from the start of the probe phase. Presumably, this signal arises from the other dot, where the resonance was lost and the nuclear field has (nearly) relaxed by the time the probe phase starts.

In order to better understand the locking mechanism, we study how far the nuclear spin polarization can be dragged by performing magnetic field sweeps as a function of the applied microwave power, the microwave frequency and the magnetic field sweep rate. Specifically, we repeatedly ramp the magnetic field from -28 mT upwards and record (i) the field at which the current jumps up (circle in Fig.~\ref{fig:sweeps}a), (ii) the field where the current jumps to a still higher value (diamond symbol), and (iii) the field where the current drops back to zero (cross). The resulting data points are shown as scatter plots in Figs.~\ref{fig:sweeps}c-e, using the same symbols. 

The first current jump always occurs as the nominal resonant field (in the absence of feedback) is first approached. The second jump and the current drop occur at fields that increase with driving amplitude over the range that we could explore (for still stronger driving, spin blockade was lifted by photon assisted tunneling, and we lost sensitivity to spin flips). For the highest powers accessible in the experiment, the electron spin is maintained on resonance over a magnetic field range of a few 100 mT. As the power is reduced, the locking effect vanishes. Furthermore, the field that can be reached before the resonance is lost, increases with excitation frequency. Earlier measurements on the same sample showed that along with the ac magnetic field an ac electric field is generated whose amplitude for a fixed power (and magnetic field amplitude) increases roughly linearly with the excitation frequency~\cite{koppens_nature}. The dependence on driving frequency can therefore also be interpreted as stronger locking for higher electric field amplitudes. Finally, we see that for higher magnetic field sweep rates the resonance is lost at lower fields.

\begin{figure}[!b]
\includegraphics[width=6cm]{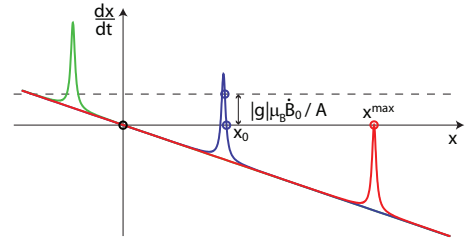}
\caption{Nuclear spin pumping curves. The nuclear spin polarization rate for one dot ($dx/dt$) is shown as a function of its polarization $x$. The overall negative slope is due to nuclear spin relaxation and the resonant peak is due to the external driving (the green, blue and red curves correspond to three different $x^\text{res}$). Circles indicate stable points in nuclear spin polarization and are found whenever the curve crosses the $x$-axis with a negative slope. During a field (or frequency) sweep, a dynamic equilibrium is reached where $dx/dt = |g|\mu_B \dot{B}_{0}/A$.}
\label{fig:pumping_curve}
\end{figure}

A few basic considerations give insight in the mechanism behind these observations. For clarity, we here present a single dot picture; the results for two coupled dots are qualitatively similar~\cite{danon_new}. We define $x$ as the dimensionless nuclear spin polarization in the dot, with $-1 \leq x \leq 1$ (in our experiments, $|x| \ll 1$).  In the absence of any excitation, the polarization naturally relaxes to zero at a rate $x/\tau_n$, due to nuclear spin diffusion. However, the nuclear spin dynamics will be affected through hyperfine-mediated electron-nuclear flip-flops when the electron spins are brought out of equilibrium~\cite{ono_prl,koppens_science,reilly_prl}. In the spin blockade regime at finite $B_0$, such non-equilibrium dynamics occurs when the electron spins are resonantly excited by an external microwave magnetic or electric field. Regardless of the relevant microscopic processes, we thus expect in very general terms a polarization-dependent pump rate $\Gamma_p(x)$, which is non-zero only close to the resonance condition $Ax^\text{res} = g\mu_B B_N^\text{res} =  |g|\mu_B B_0 - hf$. Here $x^\text{res}$ is thus the nuclear spin polarization that brings the electron spin in resonance with the excitation. The dynamics of the polarization in the dot is then described by
\begin{equation}
\frac{dx}{dt} = \Gamma_p(x) - \frac{1}{\tau_n}x.
\label{eq:pumping_curve}
\end{equation}
Fig.~\ref{fig:pumping_curve} qualitatively visualizes Eq.~\ref{eq:pumping_curve} in the form of a pumping curve for three different values of $x^\text{res}$, where we have (for now arbitrarily) chosen the resonant contribution to be positive. From the figure we can see that stable points of nuclear polarization occur when $dx/dt$ crosses zero with a negative slope: if $x$ is higher (lower) than the stable polarization $x_0$, $dx/dt$ is negative (positive) and $x$ gets pushed back to $x_0$. Due to nuclear spin relaxation, there is almost always a stable point at $x=0$. Depending on the particular shape of $\Gamma_p(x)$, hence on the specific experimental regime, there can be one or more additional stable points ~\cite{danon_new,danon_prl,rudner_reverse}.

We now interpret the field sweep experiments within this simple picture. First, given that the current remains high in field sweeps, a stable point must exist close to resonance, in agreement with our expectation of a resonant peak in $\Gamma_p$. Next, since dragging is generally observed only for $x>0$, $\Gamma_p$ must be positive, as in Fig.~\ref{fig:pumping_curve}. Finally, from the maximum nuclear field $B_N^\text{max}$ that can be achieved by dragging, we can estimate the height of $\Gamma_p$: when the maximum of the pumping peak falls below zero, i.e.\ when nuclear spin relaxation dominates the resonant pumping, the stable point at $x>0$ disappears and $B_N$ relaxes to zero (Fig.\ \ref{fig:pumping_curve}, red curve). 

During actual field sweeps, the resonance is lost at fields below $B_N^\text{max}$: since a dynamic equilibrium is reached when $dx/dt = |g| \mu_B \dot{B}_{0}/A$ instead of $dx/dt=0$, the stable operating point moves up the pumping curve (see Fig.~\ref{fig:pumping_curve}) and disappears when the sweep rate exceeds the maximum of the pumping peak. In practice we will loose the resonance even earlier, because intrinsic nuclear field fluctuations can drive the nuclear field across the maximum. We model the average switching field taking into account such fluctuations by assuming an exponential dependence of the switching rate on the ``barrier height''. The result is illustrated in Fig.~\ref{fig:sweeps}e. This combined picture captures very well the experimental observation that for higher sweep rates the resonance is more easily lost, but not at exactly the same field every time.

We next turn to the nature of the extrinsic pumping process, $\Gamma_p$. First, the stable points in the experiment generally occur for $x>0$, i.e.\ the nuclear field points against the external magnetic field. This is opposite to the usual Overhauser effect, where electron spins are excited by magnetic resonance and relax back from $\da$ to $\ua$ by flip-flopping with the nuclear spins. ``Reverse'' pumping is possible when there is an excess of $\ua$ electrons, which are excited to $\da$ by resonant electric fields, whereby the nuclear spins absorb the angular momentum~\cite{rudner_reverse,laird}. In our experiment, spin-exchange with the leads due to photon-assisted tunneling (at $10-100$ kHz) gives an excess of $\ua$ electrons, which favors reverse pumping. Second, the locking effect gets stronger, hence $\Gamma_p$ becomes larger, not only with stronger driving in general (Fig.~\ref{fig:sweeps}c), but also with stronger electric excitation by itself (higher $f$, Fig.~\ref{fig:sweeps}d). Based on these observations, we suggest that electric-field assisted electron-nuclear flip-flops combined with electron spin relaxation are mainly responsible for the resonant pumping~\cite{danon_new}.

Finally, we analyze theoretically the implications of our observations for the width of the nuclear field distribution. We define $\Gamma_\pm(x)$ as the total positive and negative nuclear spin flip rates that result from the intrinsic relaxation and resonant response combined, so $dx/dt = \frac{2}{N}(\Gamma_+ - \Gamma_-)$. We also define $\gamma(x)$ as the total rate of nuclear spin flips, $\gamma = \frac{2}{N}(\Gamma_+ + \Gamma_-)$. Using the fact that the pumping curve exhibits a resonant peak at $|x_0|\ll1$, we can then approximate the variance of the nuclear polarization distribution around $x_0$ as (see Supplementary Information)
\begin{equation}
\sigma^2 \approx \frac{1}{N} \frac{\gamma(x_0)}{\left( - \frac{\partial}{\partial x} \frac{dx}{dt} \right)|_{x_0} } \;.
\label{eq:sigma}
\end{equation}
The numerator is the local diffusion rate, and the denominator is the restoring force -- the steeper the slope of $dx/dt$, the stronger the restoring force. For the case without pumping, we have $\Gamma_\pm = N_\mp/2\tau_n$, so Eq. \ref{eq:sigma} gives us the usual result $\sigma^2 = 1/N$. For a stable point $x_0>0$ near resonance, we take as a rough estimate for the local slope the maximum of $\Gamma_p$ divided by its width. This gives $\sigma^2 \approx B_1/N B_N^\text{max}$ (see Supplementary Information). Since $B_N^\text{max}$ was several 100 mT with $B_1 < 1$ mT, these arguments imply that the nuclear field distribution was narrowed by more than a factor of 10.
Future experiments will aim at an explicit demonstration of narrowing via Ramsey-style experiments.

Narrowing of the nuclear field distribution would greatly enhance our level of control of the electron spin dynamics. Furthermore, the observed locking effect allows us to accurately set the spin resonance frequency of an electron in a quantum dot to a value determined only by the externally controlled excitation frequency. Finally, our measurements suggest that we can selectively control the ESR frequency in one of the dots, which could be exploited for independent addressing of electron spins in quantum dots that are less than 100 nm apart. 

We thank F. R. Braakman, P. C. de Groot, M. Laforest, L. R. Schreiber, G. A. Steele and S-C. Wang for help and discussions, and R. Schouten, A. van der Enden, R.G. Roeleveld and P. van Oossanen for technical support. This work is supported by the `Stichting voor Fundamenteel Onderzoek der Materie (FOM)' and the `Nederlandse Organisatie voor Wetenschappelijk Onderzoek (NWO)'.

\onecolumngrid
\newpage

\setcounter{page}{1}
\thispagestyle{empty}

\begin{center}
\textbf{{\large Supplementary Material for\\
``Locking electron spins into magnetic resonance by electron-nuclear feedback''}}\\
\bigskip
I.T.\ Vink, K.C.\ Nowack, F.H.L.\ Koppens, J.\ Danon, Yu.V.\ Nazarov, and L.M.K.\ Vandersypen\\
\textit{Kavli Institute of NanoScience, Delft University of Technology, 2628 CJ Delft, The Netherlands}
\end{center}

\section{Tuning the double dot}

The conditions for observing a pronounced electron-nuclear feedback are as follows. Qualitatively, the interdot tunnel coupling and the tunnel coupling to the outgoing lead are increased compared to the regime of Ref.~\cite{koppens_natures}. Furthermore, the potentials of the double dot are tuned such that the interdot transition occurs without energy loss: at low power, the configuration of the dot potentials is such that electrons can tunnel elastically from the left to the right dot when spin blockade is lifted. Thereby, the interdot transition is made from the $(1,1)$ singlet to the $(0,2)$ singlet, where $(m,n)$ represent the effective electron numbers on the two dots. This working point cannot be used at strong driving, since the electric field component of the excitation causes photon assisted tunneling to the $(0,2)$ triplet, thereby lifting spin blockade irrespective of the spin states of the two electrons. Instead, the double dot must be tuned such that the $(0,2)$ singlet electrochemical potential is higher than that of the $(1,1)$ singlet. This is nominally in the Coulomb blockade regime, but photon-assisted tunneling now provides the missing energy in order to make the transition from the $(1,1)$ to the $(0,2)$ singlet.\\

\section{Suppression of fluctuations}

In this section we derive an estimate for the typical magnitude of nuclear field fluctuations around a stable point close to resonance. For the sake of argument we show here the derivation for a \emph{single} quantum dot, although a similar argument holds for our double dot setup. In the double dot case, a \emph{two dimensional} Fokker-Planck equation must be considered, where stable points correspond to zeros of $\{ \partial_t x_1, \partial_t x_2 \}$ in the plane $(x_1,x_2)$. The results however, are qualitatively the same as in the single dot case.

We consider all possible configurations of the nuclear spin system in the dot as discrete points, labeled $n$, defining $n\equiv \frac{1}{2}(N_+ - N_-)$, where $N_{+(-)}$ denotes the number of nuclei with spin up(down)~\cite{footnote1}. This results in $N \equiv N_+ + N_-$ possible values for $n$, ranging from $-N/2$ to $N/2$. To investigate the stochastic properties we derive a Fokker-Planck equation for the probability distribution function $\mathcal{P}(n)$, starting from a simple master equation
\begin{equation}
\frac{\partial \mathcal{P}(n)}{\partial t} = -\mathcal{P}(n) [ \Gamma_+(n) + \Gamma_-(n) ] + \mathcal{P}(n-1) \Gamma_+(n-1) + \mathcal{P}(n+1) \Gamma_-(n+1).
\label{eq:master}
\end{equation}
In this equation $\mathcal{P}(n)$ gives the chance of finding the system in state $n$, and $\Gamma_\pm(n)$ is the rate at which the spin bath flips from the configuration $n$ to $n\pm 1$. We go over to the continuous limit, justified by the large number of nuclei $N\sim 10^6$~\cite{vankampen}, and expand all functions around $n$ up to second order. We find
\begin{equation}
\frac{\partial \mathcal{P}}{\partial t} = \frac{\partial}{\partial n} \left\{(\Gamma_--\Gamma_+)\mathcal{P} +\frac{1}{2} \frac{\partial}{\partial n} (\Gamma_-+\Gamma_+)\mathcal{P} \right\},
\label{eq:fp}
\end{equation}
a Fokker-Planck equation where all rates $\Gamma_\pm$ are still functions of $n$. Due to the large number of nuclei, the spin flip rates $\Gamma_\pm$ do not change on their full scale when increasing $n$ by only $\pm 1$ (the features of $\Gamma_\pm$ occur on the scale of the width of the resonance $\sim 1$~mT, whereas changing $n$ by $\pm 1$ corresponds to $A/N \sim 5\ \mu$T). This implies that $| \partial_n \Gamma_\pm | \ll \Gamma_\pm$, which allows us to neglect one of the cross terms resulting from the last term in (\ref{eq:fp}).

In the resulting continuity equation, the right-hand side corresponds to the derivative of a probability flux. In equilibrium this probability flux must vanish, which enables us to write down a general equilibrium solution of (\ref{eq:fp}). In terms of the bath polarization $x \equiv 2n/N$ this solution reads
\begin{equation}
\mathcal{P}(x) = \exp \left\{ \int^x N\frac{\Gamma_+-\Gamma_-}{\Gamma_++\Gamma_-} dx' \right\}.
\label{eq:sol}
\end{equation}
Maxima and minima of this distribution are found at the zeros of the derivative of the exponent. Suppose the point $x_0$ is one of these solutions corresponding to a maximum of $\mathcal{P}(x)$ (i.e.\ the second derivative in the point $x_0$ is negative). We then expand the exponent of $\mathcal{P}(x)$ up to second order around the maximum, giving a Gaussian approximation for $\mathcal{P}(x)$,
\begin{equation}
\mathcal{P}(x) \approx \exp \Bigg\{ \int^{x_0} N\frac{\Gamma_+-\Gamma_-}{\Gamma_++\Gamma_-} dx'  + \left. \frac{N}{2}\frac{\partial}{\partial x} \frac{\Gamma_+-\Gamma_-}{\Gamma_++\Gamma_-}\right|_{x_0}(x-x_0)^2\Bigg\} \equiv \alpha \exp \Bigg\{ -\frac{(x-x_0)^2}{2\sigma^2} \Bigg\},
\label{eq:appr}
\end{equation}
where $\sigma$ gives the width of the distribution. So we find that
\begin{equation}
\sigma^2 = \frac{1}{N} \left( -\frac{\partial}{\partial x} \left. \frac{\Gamma_+-\Gamma_-}{\Gamma_++\Gamma_-}\right|_{x_0} \right)^{-1} = \frac{1}{N} \left. \frac{\Gamma_++\Gamma_-}{\frac{\partial}{\partial x} (\Gamma_--\Gamma_+)}\right|_{x_0}.
\label{eq:sigmas}
\end{equation}
We now only still want to translate this expression in terms of the `pumping curve'. We use the relation $dx/dt = (2/N) (\Gamma_+-\Gamma_-)$ and define $\gamma(x) = (2/N) (\Gamma_++\Gamma_-)$. In the limit of small polarizations, i.e.\ $|x| \ll 1$, we can write 
\begin{equation}
\frac{dx}{dt} = L(x) - \gamma (x) x.
\label{eq:pump}
\end{equation}
In this notation the effect of $\Gamma_p$ (main text) is separated into two parts: (i) a polarization-dependent net spin pumping contribution, $L(x)$, and (ii) a polarization-dependent contribution to the relaxation, which together with the intrinsic relaxation rate $1/\tau_n$ is written as $\gamma (x)$. One can rewrite equation (\ref{eq:sigmas}) in terms of $dx/dt$ and $\gamma (x)$ using the relations given above. This gives us finally the expression
\begin{equation}
\sigma^2 \approx \frac{1}{N} \frac{\gamma(x_0)}{\left. \left(-\frac{\partial}{\partial x} \frac{dx}{dt}\right) \right|_{x_0}}.
\label{eq:sigma2}
\end{equation}
To get an idea of the magnitude of this variance, we approximate the derivative of the pumping curve at the stable point as roughly the height of $L(x)$ over the width (see Fig.\ 4 in the main text), i.e.\ $-\partial_x (dx/dt)|_{x_0} \approx L^\text{max} / \tilde x$, where $\tilde x$ is the width of $L(x)$. From equation (\ref{eq:pump}) we see that we can write for the absolute maximum of achievable polarization $x^\text{max} = L^\text{max} / \gamma(x^\text{max})$. Combining these two expressions and using that $\gamma (x^\text{max}) \sim \gamma(x_0)$, we find the order of magnitude of the variance $\sigma^2$ to be
\begin{equation}
\sigma^2 \sim \frac{1}{N} \frac{\tilde x}{x^\text{max}}.
\label{eq:sigma3}
\end{equation}
In terms of the effective nuclear field $B_N$, this variance reads
\begin{equation}
\sigma_{B_N}^2 \sim \Omega^2 \frac{B_1}{|B_N^\text{max}|},
\label{eq:sigma4}
\end{equation}
where $\Omega \equiv A/g\mu_B\sqrt{N}$ are the diffusive fluctuations around the unpolarized state, and $B_1$ is the scale of the width of the pumping term $L$, in our case given by the strength of the microwave driving field.

\section{Statistics of switching}

Here we explain how we calculated the purple curve in Fig.\ 4e in the main text. We suggest that the second current jump (red diamonds in the Figure) corresponds to the resonance being lost in one of the two dots. This occurs when the effective barrier between the polarized and unpolarized states becomes small enough for a typical nuclear field fluctuation to overcome. If we assume a simple linear decrease of this effective barrier for increasing $B_N$ and include the effect of the finite sweep rate $\dot B_0$, we find the polarization-dependent switching rate
\begin{equation}\label{eq:gsw}
\Gamma_\text{sw}(B_N) = \Gamma_0 \exp \left\{\gamma \left( \frac{B_N}{B_N^\text{max}} + \frac{\dot B_0}{\dot B_0^\text{max}} \right) \right\},
\end{equation}
where $\dot B_0^\text{max}$ is the maximal sweep rate to observe any locking at all. From this expression we can derive the standard deviation in $B_N$ where the second jump is observed, $\sigma_\text{sw}$, and the average switching field $\langle B_N^\text{sw} \rangle$. Explicitly, we find
\begin{equation}\label{eq:estsw}
\sigma_\text{sw} = \frac{B_N^\text{max}}{\gamma} \qquad\text{and}\qquad \langle B_N^\text{sw} \rangle = \sigma_\text{sw} \ln \frac{\dot B_0^\text{max}}{\sigma_\text{sw}\Gamma_0} + \sigma_\text{sw}\ln \frac{\dot B_0}{\dot B_0^\text{max}}
- B_N^\text{max} \frac{\dot B_0}{\dot B_0^\text{max}}.
\end{equation}
We analyzed the set of red diamonds in Fig.\ 4e in the main text. From (\ref{eq:estsw}) we expect $\sigma_\text{sw}$ to be constant in first approximation, which is indeed observed for lower sweep rates (100-400 mT/min). The decrease of $\sigma_\text{sw}$ for sweep rates above 400 mT/min could be a consequence of the average switching field lying too close to the resonance condition. Therefore we averaged the standard deviation over the first four values to find $\sigma_\text{sw} = 39$~mT. Using this value for the standard deviation, we fitted equation (\ref{eq:estsw}) to the data in Fig.\ 4e. This resulted in the fitting parameters $B_N^\text{max} = 289.6$~mT, $\dot B_0^\text{max} = 920.7$~mT/min and $\gamma = 6.946 \cdot 10^{-4}$~s$^{-1}$, giving a sample correlation coefficient of $R=0.948$: The resulting fitting curve is plotted in purple in Fig.\ 4e. Another way to estimate $B_N^\text{max}$ and $\dot B_0^\text{max}$ is to extrapolate the set of red diamonds in Fig.\ 4e to the two axes. In this way one finds the estimates $B_N^\text{max} \approx 300$~mT and $\dot B_0^\text{max} \approx 900$~mT/min, both in reasonable agreement with the results of the fit.

\section{Analysis of ESR current levels}

\begin{figure}[t]
\includegraphics[width=8.5cm]{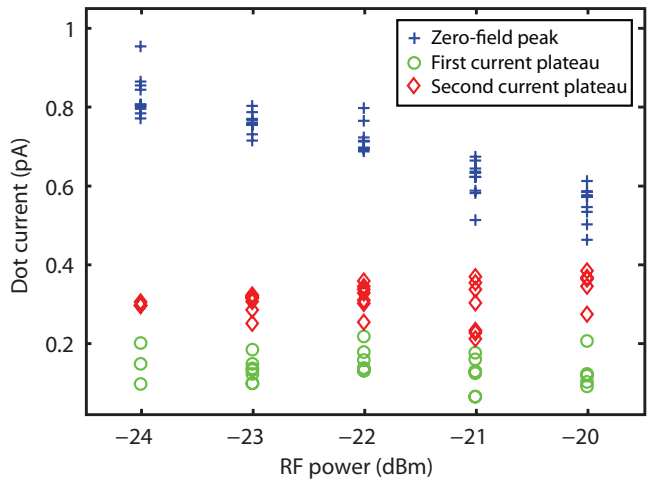}
\caption{
Current levels of the zero-field peak and the two plateaus. An offset is substracted from all current levels given by the average current between the tail of zero-field peak and the ESR resonance. The height of the zero-field peak is determined by averaging  3 points around the position of its maximum, which is determined by first averaging 10 consecutive measurements and determining the maximum current in the averaged trace. The current levels of the first plateau are obtained by averaging individual traces between the magnetic field values where the first step occurs (indicated by green circles in Fig. 4a in the main text) and the field where the second step occurs or the field value where the current drops to zero, if that occurs before the second step (red diamonds and black crosses in Fig. 4a in the main text). We require these magnetic field intervals to be longer than 10 measurement points (corresponding to 20mT) in order not to be omitted. The height of the second plateau is determined in a similar way but now by averaging between the magetic field values where the second step occurs and the field where the current drops to zero. The resulting heights of the zero-field peak, first and second current plateaus are represented here by respectively blue crosses, green circles and red diamonds for different excitation powers.
}
\label{fig:currentlevels}
\end{figure}

Next to the \emph{position} of the current jumps, we also analyzed the \emph{height} of the current plateaus between the jumps as function of driving amplitude. For different microwave powers we repeatedly swept the external magnetic field from low to high with a sweep speed of $\dot B_0=400$~mT/min, keeping the driving frequency fixed at $f=400$~MHz. For each trace we averaged the current of the first plateau and the current of the second plateau, and we determined the height of the zero-field peak.
The result is plotted in Figure \ref{fig:currentlevels} as a scatter plot for the different microwave powers. We clearly observe that in all traces the highest current was measured in the zero-field peak, and that the second plateau exhibited higher current than the first. As to the dependence of the current levels on driving power, we see that (i) the height of the zero-field peak tends to decrease with increasing excitation power and (ii) the height of the ESR current plateaus seems nearly constant. As we attribute the observed double step feature to dragging of the nuclear field, first in two and then only in one dot, we here give some general considerations concerning the current levels during resonant electron transport in double quantum dot ESR experiments.

Let us first consider the limit of strong microwave driving with a saturated ESR, i.e.\ $g\mu_BB_1/h$ much larger than all relaxation and decay rates. If \emph{both} dots are exactly on resonance, the driving causes the electrons to evolve entirely within the triplet subspace~\cite{koppens_natures}, i.e.\ in the cycle $\ket{T_+} \to \frac{1}{2}\left\{ \ket{T_+}+\sqrt{2}\ket{T_0} + \ket{T_-} \right\} \to \ket{T_-} \to  \frac{1}{2}\left\{ \ket{T_+}-\sqrt{2}\ket{T_0} + \ket{T_-} \right\} \to \ket{T_+}$.
As all three $(1,1)$ triplet states are Pauli spin blockaded, current can only flow to the extent there is relaxation from the triplets to the singlet.
If only \emph{one} of the two dots is on resonance, the system will evolve due to driving in the cycle $\ket{T_\pm} \to \frac{1}{\sqrt{2}}\left\{ \ket{T_0} \pm \ket{S}\right\} \to \ket{T_\pm}$, where in the course of every cycle the state $\frac{1}{\sqrt{2}}\left\{ \ket{T_0} \pm \ket{S}\right\}$ can decay via the $(0,2)$ singlet to the outgoing lead, giving rise to a current.
Therefore, we expect in this limit of strong driving to observe the highest current when only one dot is on resonance.
Since the resonance is saturated in the strong driving regime, we expect to first approximation no dependence of the current on microwave power.

In the limit of very weak driving, with $g\mu_BB_1/h$ much smaller than the relevant rates, one would expect quite the opposite. In this case the system spends most time in a Pauli spin blockaded state. The blockade can be lifted by spin relaxation in one of the two dots or by a spin flip in either of the dots caused by the driving field $B_1$. In this limit we therefore expect increasing current with increasing driving power, and furthermore that current will be highest when both dots are on resonance, simply because more spin flips take place.

During a field or frequency sweep, it is in principle possible that a nuclear field builds up in only one dot when the nominal ESR condition is first reached, subsequently locking the dot to the ESR condition. However, it is very unlikely that a nuclear field would build up in the other dot at a later time, when the ESR frequency in that dot is very far away from the driving frequency. A much more likely scenario is that a nuclear field builds up in both dots when the ESR condition is first reached (first plateau, low current), and that at the second current jump, the polarization in one dot relaxes to zero and only the other dot polarizes further (second plateau, high current). This would suggest that our experiments were performed in the regime of strong driving. 

However, there is an issue which does not fit in this simple picture. The decrease of the zero-field peak height for increasing power suggests that the electric field component of the excitation smears out the current peak in gate voltage space due to photon-assisted tunneling (at high frequencies, discrete sidebands are visible, at low frequencies, the sidebands overlap). This could account for the decrease of the zero-field current, but should presumably affect the ESR current levels in the same way since the ESR transition is saturated at strong driving. However, experimentally, the current levels at the two ESR plateaus are roughly independent of power, rather than decreasing with power. This point remains at present unresolved.

In order to develop a coherent picture of electron transport at zero-field and at spin resonance, a more systematic and detailed study of the dependence of the current levels on driving power and on the tuning of the double dot (tunnel coupling, detuning) is needed. This is quite involved, since the behavior of even the zero-field peak varies widely with tuning parameters.


\begin{thebibliography}{99}
\bibitem{footnote1} We assume here for simplicity nuclear spin $I=1/2$. For higher values of spin, e.g.\ $I=3/2$ as in GaAs, the results do not change qualitatively.
\end{thebibliography}
\end{document}